\documentclass{webofc}
\usepackage[varg]{txfonts}   
\usepackage{multirow}
\usepackage{graphicx}

\newcommand{\beq}{\begin{equation}}
\newcommand{\eeq}{\end{equation}}
\newcommand{\bea}{\begin{eqnarray}}
\newcommand{\eea}{\end{eqnarray}}

\newcommand{\eq}{\begin{equation}}
\newcommand{\en}{\end{equation}}
\newcommand{\eqa}{\begin{eqnarray}}
\newcommand{\ena}{\end{eqnarray}}
\newcommand{\half}{\frac{1}{2}}
\newcommand{\tr}{\mbox{Tr}}
\begin{document}
\title{Decomposition of the static potential in the
Maximal Abelian gauge}

\author{ \firstname{Vitaly} \lastname{Bornyakov}\inst{1,2,3}\fnsep\thanks{\email{vitaly.bornyakov@ihep.ru}} \and
        \firstname{Vladimir} \lastname{Goy}\inst{2,4}
\and
\firstname{Ilya} \lastname{Kudrov}\inst{3}
\and
\firstname{Roman} \lastname{Rogalyov}\inst{1}
}

\institute{Institute for High Energy Physics NRC Kurchatov Institute, 142281 Protvino, Russia
\and
         Pacific Quantum Center, Far Eastern Federal University, 690950 Vladivostok, Russia 
\and
         Institute of Theoretical and Experimental Physics NRC Kurchatov Institute, 117218 Moscow, Russia
\and
         Institut Denis Poisson CNRS/UMR 7013, Université de Tours, 37200 France
 }

\abstract{%
Decomposition of SU(2) gauge field into 
the monopole and monopoleless components 
is studied in the Maximal Abelian gauge 
using Monte-Carlo simulations in lattice
SU(2) gluodynamics as well as in 
two-color QCD with both
zero and nonzero quark chemical potential. 
The interaction potential 
between static charges
is calculated for each component 
and their sum is compared with 
the non-Abelian static potential. 
A good agreement is found 
in the confinement phase. 
Implications of this result are discussed.
}
\maketitle
\section{Introduction}
\label{intro}
We study the decomposition of the non-Abelian gauge field 
in the Maximal Abelian gauge (MAG) \cite{Kronfeld:1987ri,tHooft:1981bkw} 
into the sum of the monopole component and the monopoleless component. 
For this purpose we employ the SU(2) lattice gauge theory. 

In terms of vector potentials 
$A_\mu(x)$, the decomposition has the form
\begin{equation}
 A_\mu(x) =  A_\mu^{mod}(x) + A_\mu^{mon}(x)  
\label{eq:decomposition}
\end{equation}
where $A_\mu^{mon}(x)$ is the monopole component 
and $A_\mu^{mod}(x)$ is the monopoleless 
component defined below and referred to as the modified gauge field. 

In the MAG, the Abelian dominance for the string tension 
has long been known 
\cite{suzuki1,suzuki2,bbms,bm,Sakumichi:2014xpa} 
(for a review see e.g. \cite{Chernodub:1997ay,Haymaker:1998cw}).
Moreover, it was found \cite{suzuki3,stack,bbms} that the properly determined 
monopole component of the gauge field produces the string tension close
to its exact  value in agreement with 
conjecture that the monopole degrees of 
freedom are responsible for confinement \cite{tHooft:1975pu,Mandelstam:1974pi}.

In Refs.~\cite{miyamura,Kitahara:1998sj}
it was shown that the topological charge, chiral condensate and
effects of chiral symmetry breaking in quenched light hadron spectrum
disappear after removal of the monopole contribution 
from the relevant operators. Similar 
computations were made within the scope of the $Z_2$ projection studies 
\cite{deforcrand}. 
In particular it was shown that after removal  
of P-vortices the confinement property disappears. 
We perform a similar 
removal of monopoles. 
We consider the following types of the static potential: $V_{mod}(r)$ obtained from the Wilson 
loops of the modified gauge field $U_\mu^{mod}(x)$, 
$V_{mon}(r)$ obtained from the Wilson loops of the monopole gauge field $u_\mu^{mon}(x)$ and the sum of these two static potentials. 
 
For one value of lattice 
spacing it was shown \cite{Bornyakov:2005hf} 
that $V_{mod}(r)$ can be well approximated 
by purely Coulomb fit function and the sum $V_{mod}(r)+V_{mon}(r)$ 
provides a good approximation to the original non-Abelian static potential 
$ V(r)$ at all distances. 

Here we study this phenomenon at three lattice spacings 
using the Wilson action and thus 
we can draw conclusions about the continuum limit. 
We also present the results for one lattice spacing 
obtained with the improved lattice field action thus checking the universality. 
Furthermore, we present results for the SU(2) theory with dynamical quarks, 
i.e. for QC$_2$D.  These results were partially presented in \cite{Bornyakov:2021enf}.

\section{Details of simulations}
\label{sec-1}
We study the SU(2) lattice gauge theory. 
Vector potentials of the gauge field 
can be defined in terms of the link variables 
by the formula
\eq
A_{\mu}(x) = {1\over 2iag}\;\Big( U_{x\mu}-U_{x\mu}^{\dagger}\Big)\;,
\en
where $a$ is the lattice spacing.
Up to terms of the order $\underline{O}(a^2)$, 
the decomposition (\ref{eq:decomposition}) 
can be rearranged to the form
\eq\label{eq:decomposition_U}
 U_\mu(x) =  U_\mu^{mod}(x) u_\mu^{mon}(x)\,   
\en
which furnishes the subject of our research.

To fix the MAG we use the simulated annealing algorithm \cite{bbms} 
with one gauge copy per configuration. 
Usually after fixing MAG the following decomposition of the non-Abelian lattice gauge field $U_\mu(x)$ is made
\eq
U_\mu(x) = C_\mu(x) u_\mu(x)\quad ,
\label{coset}
\en
where $u_\mu(x)$ is  the Abelian field and $C_\mu(x)$ 
is the non-Abelian coset field. 
The Abelian gauge field $u_\mu(x)$ is further decomposed \cite{svs} into 
the  monopole (singular)
part $u^{mon}_\mu(x)$ and the photon (regular) part $u^{ph}_\mu(x)$:
 \eq
u_\mu(x) = u^{mon}_\mu(x) u^{ph}_\mu(x)\ .
\en
Then it follows from eq.~(\ref{eq:decomposition}) that
\eq
U_\mu^{mod}(x)= C_\mu(x)u^{ph}_\mu(x).
\en
Note that $u^{ph}(x)$ is the Abelian projection
of $U_\mu^{mod}(x)$ and involves no monopoles.

We need to compute the usual Wilson loops
\eq
W(C) = \half \tr {\cal{W}}(C)\,,\quad      {\cal{W}}(C) = \left(\prod_{l \in C} U(l)\right) \quad,
\label{wnonab}
\en
the monopole Wilson loops
\eq
W_{mon}(C) = \half \tr\left(\prod_{l \in C} u^{mon}(l)\right) \quad,
\label{wmon}
\en
and the non-Abelian Wilson loops with removed monopole contribution 
\eq
W_{mod}(C) = \half \tr  {\cal{W}}_{mod}(C)\,,\quad   {\cal{W}}_{mod}(C)  =   \left(\prod_{l \in C} U^{mod}(l)\right) \quad,
\label{woff}
\en
It is known that MAG fixing leaves $U(1)$ gauge symmetry unbroken. 
The monopole Wilson loop $ W_{mon}(C)$ is invariant under respective 
residual gauge transformations.
This is not true for $ W_{mod}(C)$ \cite{Bornyakov:2005hf}. 
Thus we need to fix the Landau $U(1)$ gauge 
by finding the maximum of the gauge-fixing functional, 
\eq
\max_{\omega} \sum_{x,\mu} \mathrm{Re}(\omega(x)u_\mu(x)\omega^{\dagger}(x+a\hat\mu))\,, 
\label{landau}
\en
where $\omega \in U(1)$ is the gauge transformation.
To imrove the noise-to-signal ratio for the static potential we use 
the APE smearing \cite{ape} in computations of the 
Wilson loops. 

\begin{figure}[htb]
\vspace*{-2.0cm}
\hspace*{-4cm}
\includegraphics[width=18cm,angle=0]{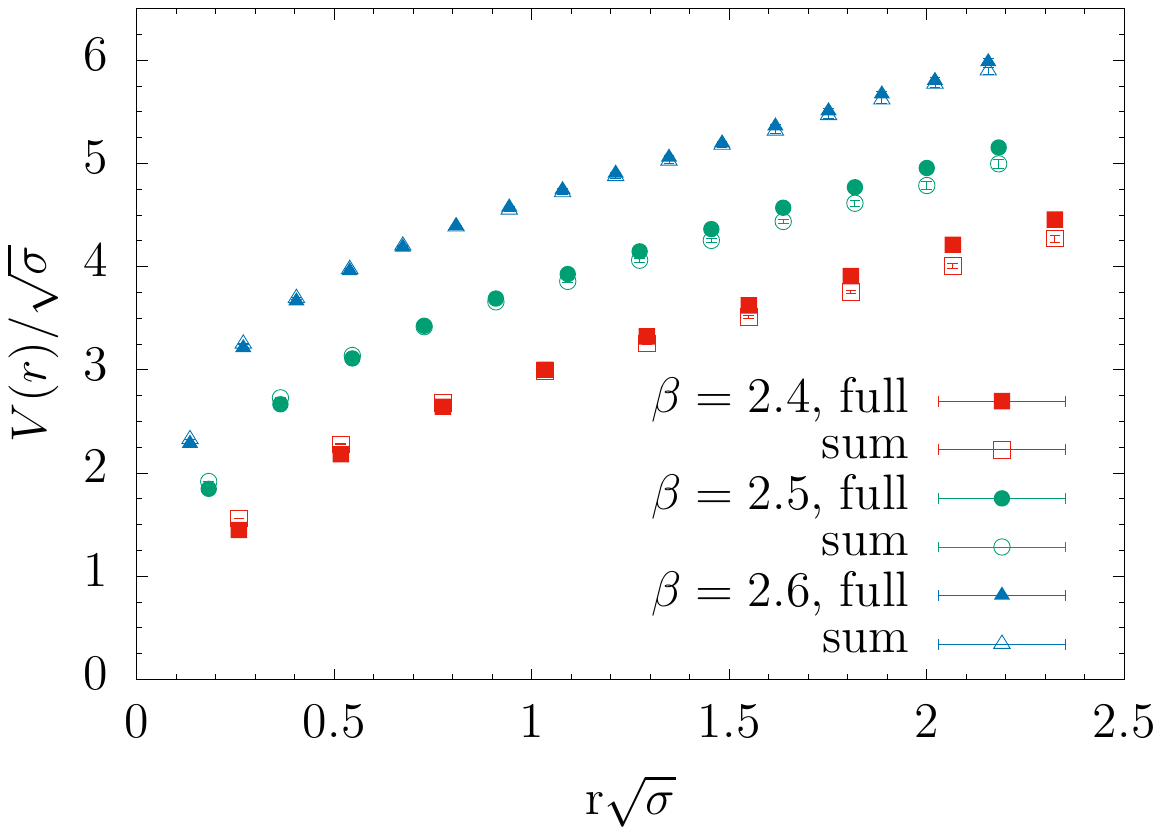}
\vspace*{-16.0cm}
\caption{Non-Abelian static potential $V(r)$  (filled symbols)
is compared with the
sum  $V_{mon}(r)+V_{mod}(r)$ (empty symbols) 
for Wilson action at $\beta=2.4$ (squares),   $\beta=2.5$ (circles),
$\beta=2.6$  (triangles). 
}
\label{potentials1}
\end{figure}

We generate 100 statistically independent 
gauge field configurations with the Wilson lattice action 
at $\beta=2.4, 2.5$ 
on the $24^4$ lattice and at $\beta=2.6$ on the $32^4$ lattice. 
The respective values of lattice spacing are 
$a=0.118,\ 0.085$ and $0.062$~fm, which are determined 
from a fit to the lattice data \cite{Fingberg:1992ju} on the string tension,
whose ``experimental'' value is set to $\sqrt{\sigma}=440$~MeV.
As a check of universality the computations were done with the tadpole 
improved action at $\beta=3.4$ on $24^4$ lattices.
Additionally we present our results \cite{Bornyakov:2017txe} obtained in QC$_2$D on $32^4$ lattice at zero and nonzero quark chemical potential $\mu_q$.

It is worth to note that another decomposition, namely eq.~(\ref{coset})
was investigated in Ref.~\cite{Sakumichi:2014xpa} in the case of SU(3)
gluodynamics. Good agreement between the static potenial $V(r)$ and the
sum $V_{abel}(r)+V_{off}(r)$ was found. We believe that this decomposition
also deserves further study.

\section{Results}
\label{sec-2}

 \begin{figure}[htb]
\vspace*{-2cm}
 \hspace*{-40mm}
  \includegraphics[width=18cm,angle=0]{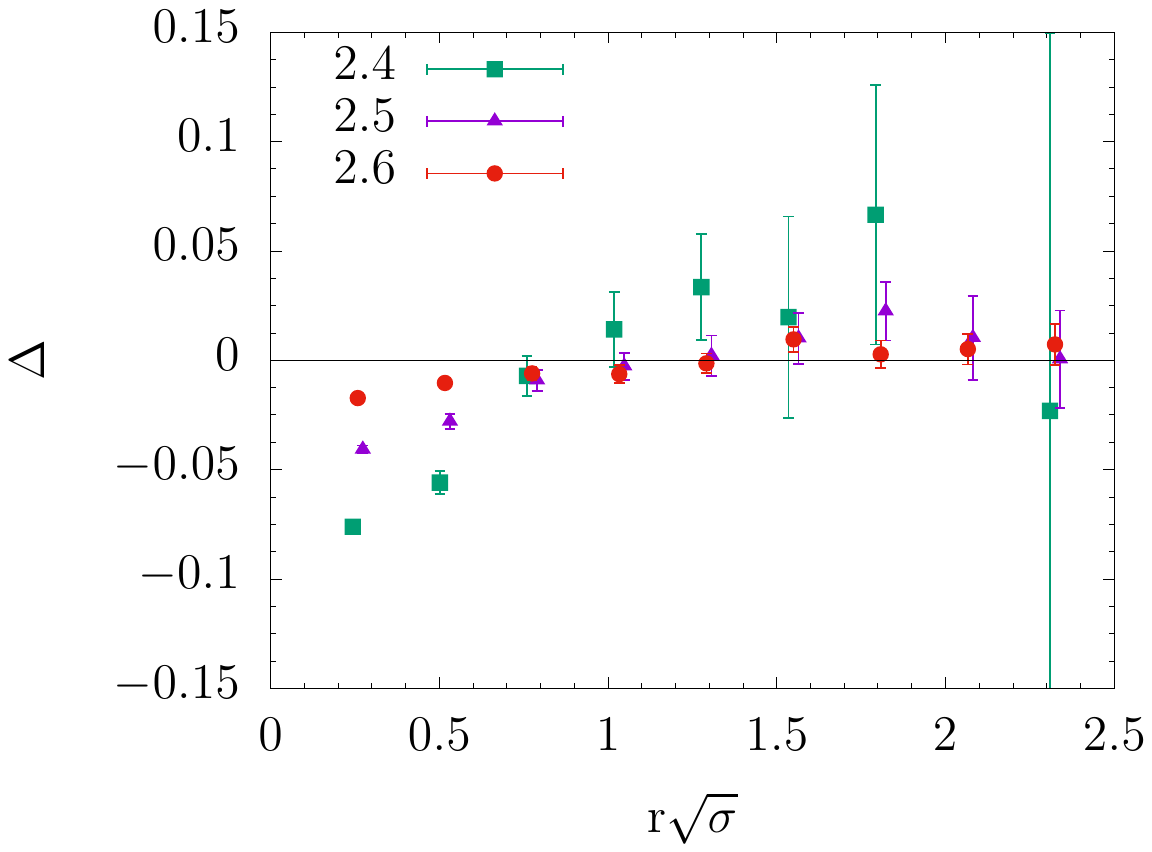}
  \vspace*{-17.0cm}
 \caption{The relative deviation $\Delta(r)$ vs. distance $r$ for three 
 values of the lattice spacing.}
 \label{delta}
 \end{figure}
\begin{figure}[htb]
\vspace*{-20mm}
\hspace*{-3cm}
 \includegraphics[width=18cm,angle=0]{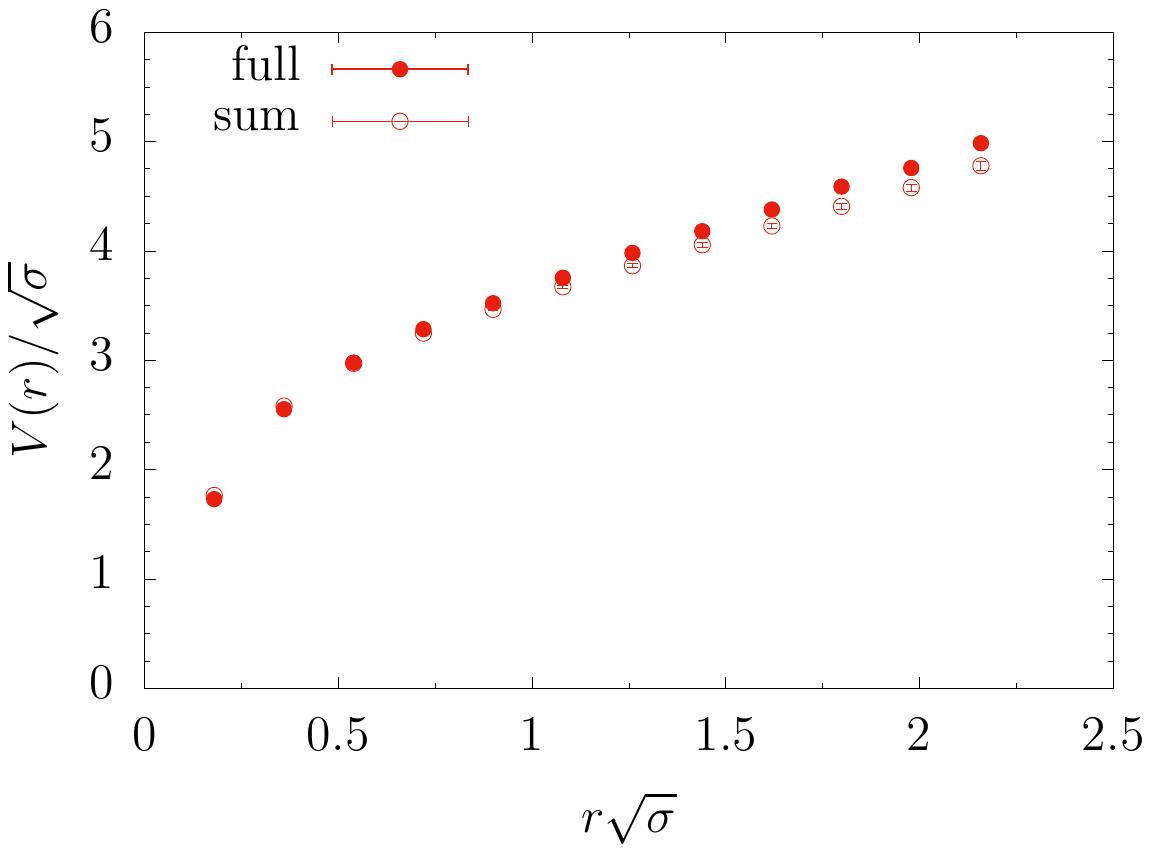}
\vspace*{-17.0cm}
\caption{Non-Abelian static potential $V(r)$  (filled symbols)
is compared with the sum of the monopole and modified potentials $V_{sum}(r)$ 
(empty symbols) for the improved action at $\beta=3.4$.} 
\label{potentials2}
\end{figure}
In Fig.~\ref{potentials1} we show the usual non-Abelian static potential 
$V(r)$ denoted as 'full' and compare it with the sum 
$V_{sum}(r)=V_{mon}(r)+V_{mod}(r)$.
One can see that approximate equality
\eq
V(r) \approx V_{sum}(r)  
\label{poten_decomp}
\en
is satisfied for all 
three lattice spacings and the approximation improves toward the 
continuum limit.
To give further support to this statement we plot in Fig.~\ref{delta} 
the relative deviation $\Delta(r)$ defined as
\beq
\Delta(r) = \frac{V(r) -  V_{sum}(r)}{V(r)}\;.
\label{eq:reldev}
\eeq
It is clear that $\Delta(r)$ decreases with decreasing lattice spacing.

\begin{figure}[htb]
\vspace*{-2.0cm}
\hspace*{-3cm}
\includegraphics[width=18cm,angle=0]{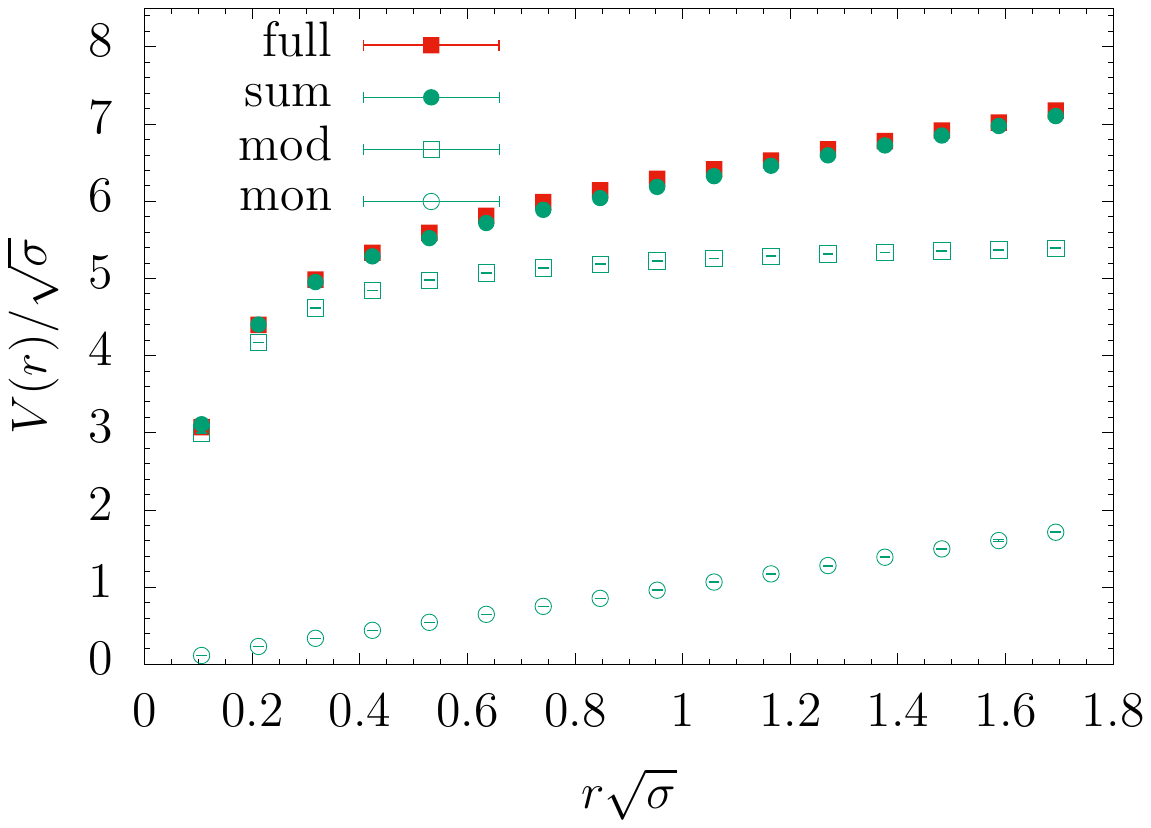}
\vspace*{-17.0cm}
\caption{The non-Abelian static potential $V(r)$  (filled squares)
is compared with the sum  of the monopole and modified potentials $V_{sum}(r)$ 
(filled circles) for QC$_2$D at zero chemical potential. The potentials 
$V_{mon}(r)$ (empty circles) and $V_{mod}(r)$ (empty squares) are also shown.} 
\label{potentials3}
\end{figure}
\begin{figure}[htb]
\vspace*{-2.0cm}
\hspace*{-3cm}
\includegraphics[width=18cm,angle=0]{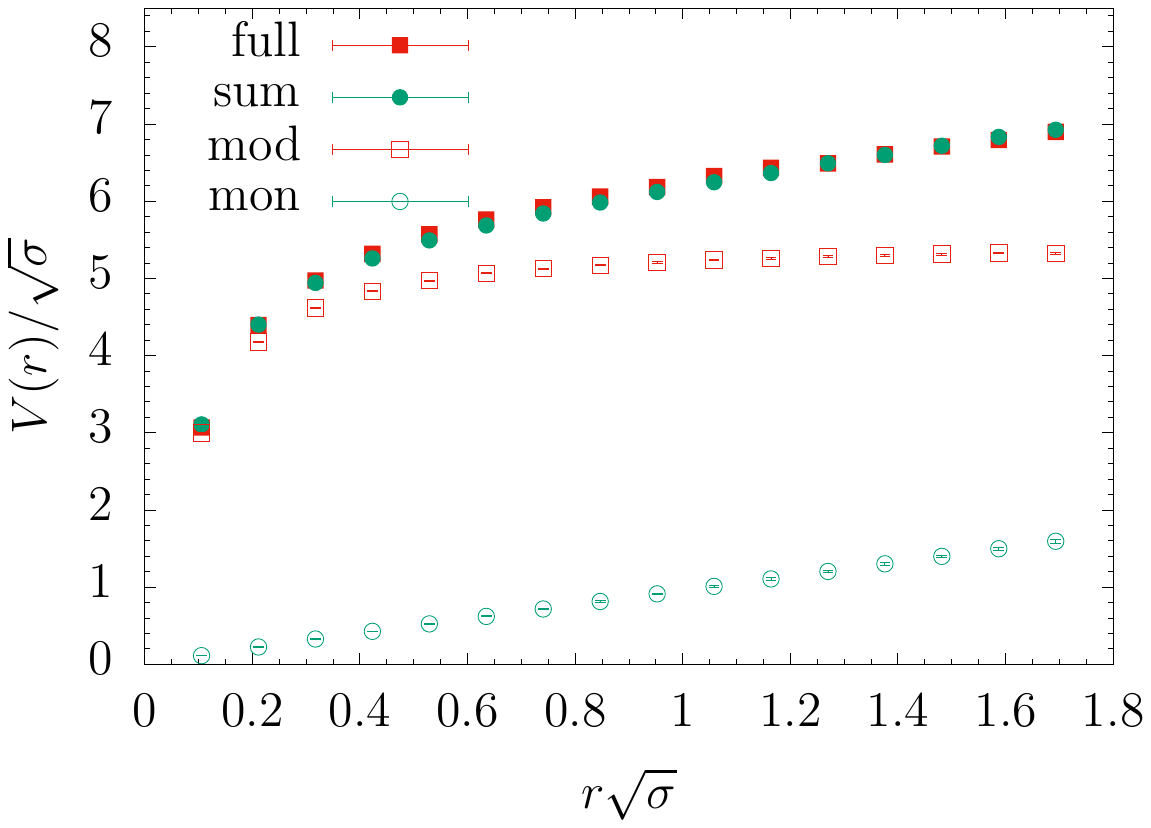}
\vspace*{-17.0cm}
\caption{The non-Abelian static potential $V(r)$  (filled squares)
is compared with the sum  of the monopole and modified potentials $V_{sum}(r)$ 
(filled circles) in QC$_2$D at $a\mu_q = 0.19$. 
The potentials 
$V_{mon}(r)$ (empty circles) and $V_{mod}(r)$ (empty squares) are also shown.} 
\label{potentials4}
\end{figure}
We also studied the universality of the decomposition of
the static potential eq.~(\ref{poten_decomp}). 
The simulations were made with 
the tadpole improved action at $\beta=3.4$ with lattice spacing 
approximately equal to that of the Wilson action at $\beta=2.5$.
The results of our computations are presented in 
Fig.~\ref{potentials2}. It is seen that the agreement between $V(r)$ and 
$V_{sum}(r)$ is nearly as good as in Fig.~\ref{potentials1} for $\beta=2.5$. 

Furthermore, we completed the same computations in QC$_2$D on $32^4$ lattices 
with small lattice spacing at zero and nonzero quark chemical potential $\mu_q$
(details of simulations in QC$_2$D can be found, e.g., in \cite{Bornyakov:2017txe}). 
The results of these computations are presented in Fig.~\ref{potentials3}
for $\mu_q=0$ and in Fig.~\ref{potentials4} for $a\mu_q=0.19$.
It can be seen clearly that approximate decomposition (\ref{poten_decomp})
is fulfilled with rather high both at zero and nonzero $\mu_q$.

\section{Conclusions}
\label{sec-3}

We have studied the decomposition of the static potential into the linear
term produced by the monopole (Abelian) gauge field $U_{mon}(x)$ and the 
Coulomb term produced by the monopoleless
non-Abelian gauge field $U_{mod}(x)$. 
In the case of Wilson action we have presented the results 
(Figs.~1 and~2) for three values of lattice spacing 
to demonstrate that the agreement becomes better with
a decrease of lattice spacing. Thus our results imply that the relation 
(\ref{eq:decomposition}) becomes exact in the continuum limit.
Further work is needed to provide more evidence for this conclusion. 
Next, we have demonstrated that the decomposition (\ref{eq:decomposition})
holds true also in the case of improved lattice action (see Fig.~3). 
Furthermore, we have found that it works also 
in QC$_2$D for both zero and nonzero quark chemical potential.
It should be noticed that in Ref.~\cite{Bornyakov:2021enf} 
we also presented results for the static potential in the adjoint representation 
and found that the decomposition (\ref{eq:decomposition}) 
works quite well in this case although not so well
as in the case of the fundamental representation.
It is of course interesting to verify how decomposition (\ref{eq:decomposition})
works for other observables, e.g. action density and energy density
of the flux tube. Also the decomposition should be checked 
 in the case of SU(3) gauge group.

One can draw the following conclusions from the decomposition 
(\ref{eq:decomposition}). The monopole part of the gauge field $U_{mon}(x)$
is responsible for the classical part of the energy of the hadronic string,
whereas the monopoleless part $U_{mod}(x)$ is associated with the
fluctuating part of its energy, i.e. $U_{mod}(x)$
reproduces perturbative results at small distances and 
contributes to the nonperturbative physics at  
large distances.

{\bf Acknowledgments.}
Computer simulations were performed on the Central Linux Cluster of the NRC “Kurchatov Institute” - IHEP
(Protvino) and Linux Cluster of the NRC “Kurchatov Institute” - ITEP (Moscow). This work was supported by the Russian Foundation for Basic Research, grant no.20-02-00737 A. The authors are grateful to G. Schierholz, T. Suzuki,
S. Syritsyn, V. Braguta, A. Nikolaev for participation at the early stages of this work and for useful discussions.

\end{document}